\documentclass[12pt,preprint]{aastex}

\slugcomment{To appear in ApJ Letters}

\shorttitle{The open cluster Trumpler 20}
\shortauthors{Platais et al.}

\begin{document}


\title{The effects of differential reddening and stellar rotation on
the appearance of multiple populations in star clusters:
the case of Trumpler 20}


\author{I. Platais\footnote{This is WOCS paper 49 of the WIYN Open Cluster
 Study.}}
\affil{Department of Physics and Astronomy, The Johns Hopkins University,
Baltimore, MD 21218, USA;}
\email{imants@pha.jhu.edu}

\author{C. Melo}
\affil{European Southern Observatory,  Karl-Schwarzschild-Str. 2, 85748 Garching
bei M\"{u}nchen, Germany}

\author{S. N. Quinn\footnote{Now at Department of Physics and Astronomy, Georgia State University, Atlanta, GA 30302, USA.}}
\affil{Harvard-Smithsonian Center for Astrophysics, 60 Garden Street, Cambridge, MA 02138, USA}

\author{J. L. Clem}
\affil{Department of Physics and Astronomy, Louisiana State University,
Baton Rouge, LA 70803, USA}

\author{S. E. de Mink\footnote{Hubble Fellow.}}
\affil{Space Telescope Science Institute, 3700 San Martin Drive, Baltimore,
MD 21218, USA}

\author{A. Dotter}
\affil{Space Telescope Science Institute, 3700 San Martin Drive, Baltimore,
MD 21218, USA}

\author{V. Kozhurina-Platais}
\affil{Space Telescope Science Institute, 3700 San Martin Drive, Baltimore,
MD 21218, USA}

\author{D. W. Latham}
\affil{Harvard-Smithsonian Center for Astrophysics, 60 Garden Street, Cambridge, MA 02138, USA}

\and

\author{A. Bellini}
\affil{Dipartimento di Astronomia, Universit\`{a} di Padova, Vicolo dell'’Osservatorio 3, Padova, I-35122, Italy}


\begin{abstract}
We present a detailed analysis of the upper main sequence of the
$\sim$1.3-Gyr old open cluster Trumpler~20. High accuracy {\mbox{$BV\!I$}}
photometry combined with the 
Very Large Telescope/FLAMES medium-resolution spectroscopy of 954 stars
is essential to understanding the unusual appearance of
the color--magnitude diagram (CMD), initially suggesting multiple populations
in Trumpler~20.
We show that differential reddening is a dominant contributor to the
apparent splitting/widening of the main-sequence turnoff region.
At its extreme, the excess differential reddening reaches $\Delta(B-V)\sim0.1$
while the adopted minimum reddening for the cluster is $E(B-V)=0.36$.
A unique sample of measured projected rotational velocities indicates
that stellar rotation is high near the main-sequence turnoff,
reaching $v\sin i \sim$180 km~s$^{-1}$. 
By dividing the upper main-sequence stars into equal groups of slow
and fast rotators, we find that fast rotators have a marginal blueshift
of $\delta(V-I)\sim-0.01$, corresponding to a difference in
the median $v\sin i$ of $\sim$60 km~s$^{-1}$ between these subsamples.
We conclude that stellar rotation has an insignificant effect on
the morphology of the upper main sequence of this intermediate-age
open cluster.
Trumpler~20 appears to contain a single coeval population of stars
but there is evidence that the red clump is extended.
\end{abstract}



\section{Introduction}

A $\sim$1.3-Gyr old Galactic open cluster, 
Trumpler~20 (hereafter Tr~20;
$\ell=301\fdg5$ $b=+2\fdg2$) was first correctly characterized by
\citet{pla08}. Its basic properties such as age, distance, and
reddening were later confirmed by \citet{sel10} and \citet{car10}.  
These studies also
indicate that the red clump of Tr~20 has a peculiar morphology; possibly,
it is dual. In addition, the upper main sequence appears to show
an enlarged color spread that is not normally seen in the color--magnitude
diagram (CMD) for open clusters in this age group. This observational
evidence hints at the now-popular conjecture of multiple stellar populations,
suggested for a number of the Magellanic Cloud (MC) clusters
\citep[e.g.,][]{mac08,mil09,gou09}. Recent studies
of rich intermediate-age (1--2~Gyr) star clusters in the Large
Magellanic Cloud (LMC) indicate that their extended main-sequence
turnoff (MSTO) regions can be described by continuous star
formation, lasting $\sim$0.3 Gyr or longer \citep{gou11,rub11}.
This long period of star formation, however, seems to be at odds with the fact that
none of the younger clusters are known to have such a trait.
Alternative interpretations include, for example, stellar rotation effects
\citep{bas09} and interactions in binary systems \citep{yan11}.
However, according to \citet{gir11}, stellar rotation is not a likely
physical process to produce extended MSTO.
Although extended star formation is generally accepted as the main
factor in generating atypical CMDs for a number of MC star clusters,
this scenario is based solely on interpretation of the photometric data.
An exception is the analysis of chemical abundances in four
intermediate-age LMC clusters \citep{muc08}. Among them, NGC~1783 is
a star cluster with an extended MSTO \citep{gou11}; however, it lacks
convincing anticorrelation in the Na--O and Mg--Al abundances, observed in the
Galactic globular clusters with multiple populations \citep[e.g.,][]{gra11}.
This might be an indication that younger star clusters with extended MSTO
may not experience self-enrichment.

The open cluster  Tr~20 is important for two reasons: (1) if proven, it
would be the first known {\it open cluster} with multiple stellar
populations in the Milky Way;  (2) the relative proximity of Tr~20 ($d=3.3$~kpc)
permits high-resolution spectroscopy of its upper main sequence. This
provides the crucial observational data on projected rotational
velocities, $v \sin i$, which are scarce for the main-sequence stars
of 1--2~Gyr old star clusters. Such data can be used
to test theoretical predictions on the significance of stellar
rotation on morphology of CMD. 
In this Letter, we discuss the results of high-accuracy photometry
of Tr~20 combined with measured radial velocities and $v \sin i$ for the
entire upper part of the cluster's CMD.
Detailed results of this study will be published elsewhere
(I.~Platais et al., in preparation).

\section{Observational data}

\subsection{{\mbox{$BV\!I$}} Photometry}

The photometric {\mbox{$BV\!I$}} CCD observations of Tr~20 were obtained
on three
observing runs in 2008--2009 at the CTIO 1.0~m telescope, operated
by the Small and Moderate Aperture Research Telescope System (SMARTS)
consortium. The instrumental setup included an STA $4064\times4064$
CCD featuring 15~$\mu$m pixels and covering a field of view of
$20\arcmin\times20\arcmin$. Exposure times for Tr~20 ranged from
10 to 600~s. Each night a large number of Landolt's standards (57--112)
were frequently observed in order to assure the best possible calibration
of instrumental magnitudes based upon empirical and field-variable
point-spread function.
Fringing in the sky background was substantial in the $I$ filter
and was removed accordingly. The observed spread of extinction, color,
and zeropoint terms in the transformations to the standard system
indicates excellent photometric quality of the sky. The combined
{\mbox{$BV\!I$}} photometry for 96,156 stars covers a $43\arcmin\times43\arcmin$
area down to $V\sim22$ with formal accuracies reaching 0.003~mag
for optimally exposed stars.

\subsection{VLT/FLAMES/GIRAFFE Spectroscopy}

Spectroscopic observations\footnote{Based on observations collected
at Paranal Observatory, ESO, program ID 083.D-0671(A).}
of Tr~20 were conducted in 2009 April--May
using the Very Large Telescope (VLT)/FLAMES/GIRAFFE multi-object spectrograph
in its MEDUSA mode,
which can allocate up to 132 fibers over a circular field-of-view
with a diameter of 25$\arcmin$. The chosen high-resolution mode
H525.8B offers resolution $R\sim25,900$ at the central
wavelength of $\lambda=$~5258~\AA. Our sample of targets contains
photometric cluster members selected from the cluster's CMD over
a field of $20\arcmin\times20\arcmin$ from \citet{pla08}. It includes
the entire red giant branch (RGB), all possible subgiants, and the
complete upper main sequence down to $V=17.0$, equivalent to
$M_V\sim+3.3$ (F2--F5\,V spectral type). The CMD of the observed stars is
discussed in Section~3. A total of 954 stars were observed using nine
fiber configurations, each taken with 1~hr of exposure time that
resulted in signal-to-noise ratio (S/N) of 15--70 for the entire sample.
Each target star was observed only once. Our goal was to maximize the number
of main-sequence cluster members, while ignoring potential binaries and
concentrating on likely single stars with  radial velocities consistent
with cluster membership.

The data were reduced using the GIRAFFE BaseLine Data Reduction
Software \citep{ble00}, which follows through the
usual steps of spectroscopic reductions. In addition,
we modeled the sky and subtracted it from the spectra of
main-sequence stars.

\subsubsection{Cross-correlating with Empirical Templates}

Radial velocities $V_{\rm r}$ and projected rotational velocities
$v \sin i$ were first derived by cross-correlating the spectrum
of each star with box-shaped-line digital templates, constructed
from the high-resolution and high-S/N spectra  (F0\,V and K0\,III, accordingly)
of slowly rotating stars \citep[e.g.,][]{bar96}. The cross-correlation
function (CCF) was then fitted with a Gaussian whose parameters
served in deriving $V_{\rm r}$ and $v \sin i$ \citep{mel01}.
These techniques work well for RGB stars, which have intrinsically
low $v \sin i$ values, and for the main-sequence stars with
$v \sin i\la 80$~km~s$^{-1}$. There is a hint that with  increasing
$v \sin i$ the radial velocities become biased, and for
$v \sin i\ga 100$~km~s$^{-1}$ this method often fails because
the narrow noise peaks, significantly enhanced by relatively low S/N
of our spectra, start to dominate the CCF \citep[e.g.,][]{gal05}.
These shortcomings of the fixed-template techniques and the desire to
measure higher $v \sin i$ cases prompted us to undertake a
different approach.

\subsubsection{Cross-correlating with Synthetic Spectra}  

To determine stellar atmospheric parameters and radial velocities, we
used the Harvard-Smithsonian Center for Astrophysics (CfA) library of
synthetic spectra, which covers the entire wavelength range of the
GIRAFFE spectra. The present version of the library was calculated by
John Laird for a grid of Kurucz model atmospheres \citep{kur05}, using
a line list developed by Jon Morse. We broadened the library to match
the actual instrumental resolution of the observed spectra, which we
determined to be FWHM=$13.3~{\rm km~s}^{-1}$ by fitting the
width of ThAr lines. Similar to the analysis in \citet{car11},
we cross-correlated each star against the grid
of synthetic spectra with the IRAF task XCSAO and, using the normalized
CCF peak heights at each grid point, we interpolated to the stellar
parameters yielding the maximum correlation value.
We assign errors of
half the grid spacing --- $125$~K in $T_{\rm eff}$, $0.5$ -- $10~{\rm
km~s}^{-1}$ in $v \sin i$ (with smaller errors for slowly
rotating stars), and $0.25$ dex in $\log g$ and [$m$/H]. Spectroscopic
determinations of $T_{\rm eff}$, $\log g$, and [$m$/H] can be
correlated, potentially causing systematic errors larger than the
nominal errors quoted here, but constraining [$m$/H] to equal the
cluster metallicity can help to break that degeneracy. We used the
red clump cluster members (see Section~3) to determine
[$m$/H]$=-0.08$ and then performed the interpolation again with the
metallicity fixed. We also note, as hinted with the other techniques,
that the radial velocities are redshifted in the amount of about
$+0.02~{\rm km~s}^{-1}$ per unit $v \sin i$ value.

\section{Cluster membership and color--magnitude diagram}

The CMD of Tr~20 shows a heavy contamination
by field stars, mostly from the Carina spiral arm \citep{sel10,car10,pla08}. 
Our sample of nearly a thousand photometrically selected probable cluster
members can now be cleaned up by using radial velocities.
First, we examined radial velocities (see Section~2.2.1)
of the red clump stars to derive the mean cluster velocity.
A sample of 68 red clump stars yields $\langle V_{\rm r}\rangle =-40.60\pm0.12$
km~s$^{-1}$ and is our adopted mean radial velocity for Tr~20.
A similar estimate from radial velocities based upon
synthetic templates gives $\langle V_{\rm r}\rangle =-40.67\pm0.12$ km~s$^{-1}$.
Second, we used $v \sin i$ and $V_{\rm r}$ to select
the probable kinematic cluster members (Figure~1).
Our final sample of kinematic members of Tr~20 consists of 471 stars with
a caveat that among them $\sim$50--100 stars (10\%--20\%), estimated by two
different statistical methods (ad hoc analysis of $v \sin i$ distribution
and a formal cluster membership calculation in the bins of $v \sin i$),
can still be field stars, predominantly at $v \sin i < 70$ km~s$^{-1}$. 

The CMD of our kinematic cluster-member sample (Figure~2) shows some
intriguing features. In order to represent and interpret them, here we
chose a $V,V-I$ CMD. First, the MSTO region appears to be split
into two parts: a narrow but sparse blue sequence and a bulky redder one.
We note that a split of the upper main sequence in Tr~20 is unusual
compared to the CMDs of selected LMC star clusters
\citep[e.g.,][]{gou11}, which all show a smooth and progressively
widening MSTO toward the higher luminosities.  Second, in the sense
of a sequence, the RGB stars are difficult to identify.
Essentially, there are only two distinct ``blobs'' of red stars.
The most conspicuous one at $V\sim 14.7$ represents the red clump,
which appears to be significantly elongated along both axes.
The second, at $V\sim 13.8$, represents the RGB clump, which is
well matched by the isochrones (see Figure~4).
Prior to interpreting these features, we first probed
the cluster for signs of differential reddening (DR).

\subsection{Differential Reddening}

In CMDs of star clusters older than $\sim$1~Gyr, DR
affects mostly the morphology of the MSTO region and the RGB
including the red clump \citep[e.g.,][]{pla11}.  Our approach in estimating DR
is based on the assumption that there is an empirical blue envelope of
the main sequence, indicating cluster stars with the {\it lowest} reddening
as shown by the curve in Figure~2. This envelope conforms with the adopted
youngest isochrone to the cluster's CMD. Next, we partially de-reddened each
main-sequence star (for Tr~20, down to $V\sim 17$) so that they lie exactly
on the blue envelope. A zero DR is adopted for a few probable cluster
members bluer than this envelope. Thus, in our notation, DR is never
negative.  To mitigate the potential correlation between DR and
the effects of stellar rotation, we used only the stars with
$v \sin i<95$ km~s$^{-1}$ (a total of 237). A raw map of estimated DR
(Figure~3) shows a distinct window of lower reddening in the
NW corner of Tr~20, consistent with the Galactic reddening maps
of \citet{sch98}. The amount of DR is then smoothed on a grid of tangential
coordinates at $1\arcmin$ steps (Figure~3). At each grid point, the median
DR is calculated from the nearest few measurements. The average
size of a constant reddening patch ($d\sim2\farcm5$, equivalent
to the eight nearest measures) was estimated by minimizing the
color dispersion in the DR-corrected upper main sequence.
After correcting for DR, there is
a substantial improvement in the tightness of all the features in the CMD
(Figure~2). DR is clearly the main reason for
the unusual appearance of the observed  CMD for Tr~20.
The following isochrone fits to our photometric data indicate
that the lowest reddening toward
Tr~20 is $E(B-V)=0.36$ for stars with $(B-V)_0=+0.3$, while the
maximum amount of DR reaches $\sim$0.1~mag. This range of $E(B-V)$ is consistent
with the earlier estimates of reddening \citep{pla08,car10}. 
We note that the other effects such as unresolved or interacting
binaries and rotation of stars may partially mimic the appearance of DR.

\subsection{Isochrone Fitting}

We used a set of Padova isochrones \citep{mar08} generated for
the metal abundance of $Z=0.015$. The isochrones were reddened using
the prescription of deriving a variable $E(B-V)$ given in 
\citet{fer63} and \citet{pla08}.
In order to achieve a reasonable match of
isochrones with the features of RGB, the parameter $\eta$ defined
as $\eta$=$E_{B-V}$(Sp~T)$/E_{B-V}$(B0) \citep{fer63} should be revised
to $\eta=0.945--0.17(B-V)_{0}$. Reddening in $BV$ bandpasses was translated
to $VI$ using the coefficients from \citet{sch98}. 
Our best fit is achieved with a 1.3~Gyr isochrone and the true distance modulus
$V_{0}-M_{v}$=12.60 mag. The morphology of the DR-corrected red clump is
unusual --- it is substantially extended along the magnitude axis 
($\Delta V \sim 0.5$), having its highest density at $V=14.4$. There
is a certain similarity to a double red clump of the SMC star cluster
NGC~419; however, the lack of multiple populations in Tr~20 and, hence,
the absence of the required $\sim$0.1~Gyr span of ages appears to weaken
the hypothesis of non-degenerate He cores \citep{gir09}.

We also used the isochrones from the Dartmouth Stellar Evolution
Database \citep{dot08}. A noted 0.2~Gyr age difference from the fit with
Padova isochrones (Figure~4) is likely due to 
differences in the treatment of convective
core overshoot and updated nuclear reaction rates, especially for the CNO cycle.

\section{Stellar rotation}

\citet{rox65} were first to predict the effect of stellar rotation
on the color and magnitude of a star, claiming that
stellar rotation is essentially tantamount to reddening. 
Recently, \citet{bas09} suggested that stellar rotation in stars
with masses in the range 1.2--1.7$M_\sun$ can mimic the effect of multiple
populations in star clusters by shifting these stars to redder
colors in the CMD. This proposition was contested by
\citet{gir11} on the grounds of evolutionary tracks for non-rotating
and rotating stars. These authors conclude that stellar models with
rotation appear to produce a modest blueshift in the CMD.
Clearly, our data can provide a crucial test of these predictions. 

First, from the DR-corrected $BV$ CMD we selected a sample of 168 upper
main-sequence stars down to $V\sim16.1$ --- the part of a CMD containing
a relatively high fraction of fast rotators. 
Second, the $V\!I$ CMD of this new sample
was rotated around the point $V=16.1$, $V-I=0.75$ in such a way that
the magnitude axis becomes approximately parallel to the selected short
segment of a 1.3~Gyr isochrone. The rms of the color distribution for
the modified DR-corrected $V-I$ of these stars is 0.04 mag.
The modified colors served as an argument in constructing the histograms of 
fast and slow rotators, divided into equal parts
at  $v \sin i = 90.5$ km~s$^{-1}$. 
Similar to \citet{gou09}, we used
the nonparametric Epanechnikov-kernel probability density function
to smooth the histograms, normalized relative to the maximum counts of
slow rotators.  The resulting color distributions 
are shown for two cases: using $V\!I$ photometry ``as is''
and correcting it for DR (Figure~5). Our data show that at
1.3~Gyr and in the mass range of 1.5-1.8$M_{\sun}$  stellar rotation
produces a slight to zero blueshift of stars in the DR-corrected CMD:
$\delta(B-V)=0.0$, $\delta(V-I)=-0.01$, and $\delta(B-I)=-0.01$.
All median color differences are between the fast and slow rotators. 
This effect appears to be 1.3 to 3 times magnified in the
DR-uncorrected $V\!I$ and $BI$ CMDs, accordingly, which can be
attributed to a likely anticorrelation between the rotational
blueshift and DR. 

To validate these findings, we explored the likelihood of a chance
blueshift in our sample.  We constructed a synthetic cluster with all
physical parameters close to those of our selection of Tr~20 members, 
particularly, the total number of stars on the main sequence and
the pattern of DR. The only variable was a level of reddening
``noise'' added to the grid of median DR (Figure~3). From a 1000 realizations
of a synthetic cluster, each time we randomly chose 168 stars and calculated
the median $V-I$ color difference between the ``fast'' and ``slow'' rotators.
In the absence of stellar rotation, the standard deviation
of these median differences is $\varepsilon (V-I)=0.005$ and it is only
weakly dependent on the level of the applied DR-noise.  Thus, in
a statistical sense the estimated color offset due to the stellar rotation is
a $\sim2\sigma$ result and agrees qualitatively with the theoretical
predictions by \citet{gir11}.

Near the MSTO, a $\delta(V-I)=-0.01$ color offset for the 1.3~Gyr isochrone
corresponds to an age shift of $\sim$20~Myr. Considering the median
difference of $\sim$60 km~s$^{-1}$ between the slow and fast rotators
and extrapolating it to the entire range of $v \sin i$ (180 km~s$^{-1}$),
we obtain a tentative ``dispersion'' in isochrone ages of $\sim$60~Myr.
This is only $\sim$15\% of the deduced average spread of ages reported by
\citet{gou11}.
We note that contamination of the slow-rotator sample by field
stars and fast rotators at high inclination angles masquerading as
slow rotators, can reduce the true color dispersion by $\sim$0.01 mag.
Even accounting for this, the stellar rotation appears to have
a marginal effect on the morphology of MSTO.

\section{Conclusions}

This study of the Galactic open cluster Tr~20 provides strong
evidence that DR can mimic the appearance of multiple
populations in star clusters. Despite some earlier indications, 
Tr~20 appears to contain a single coeval population of stars.
Positions of the upper main-sequence stars in the CMD of Tr~20 are
marginally dependent on their projected rotational velocities and
in a specific way. Our data show that high $v \sin i$ stars are
slightly blueshifted. Stellar rotation appears to play merely
a minor role in the observed broadening of MSTO of this
intermediate-age open cluster.

\acknowledgments

We thank Nate Bastian and the anonymous referee for meritorious
suggestions.  This work has been supported in part by the NSF Grant AST
09-08114 to JHU (I.P.).

\clearpage

\clearpage

\begin{figure}
\epsscale{.90}
\plotone{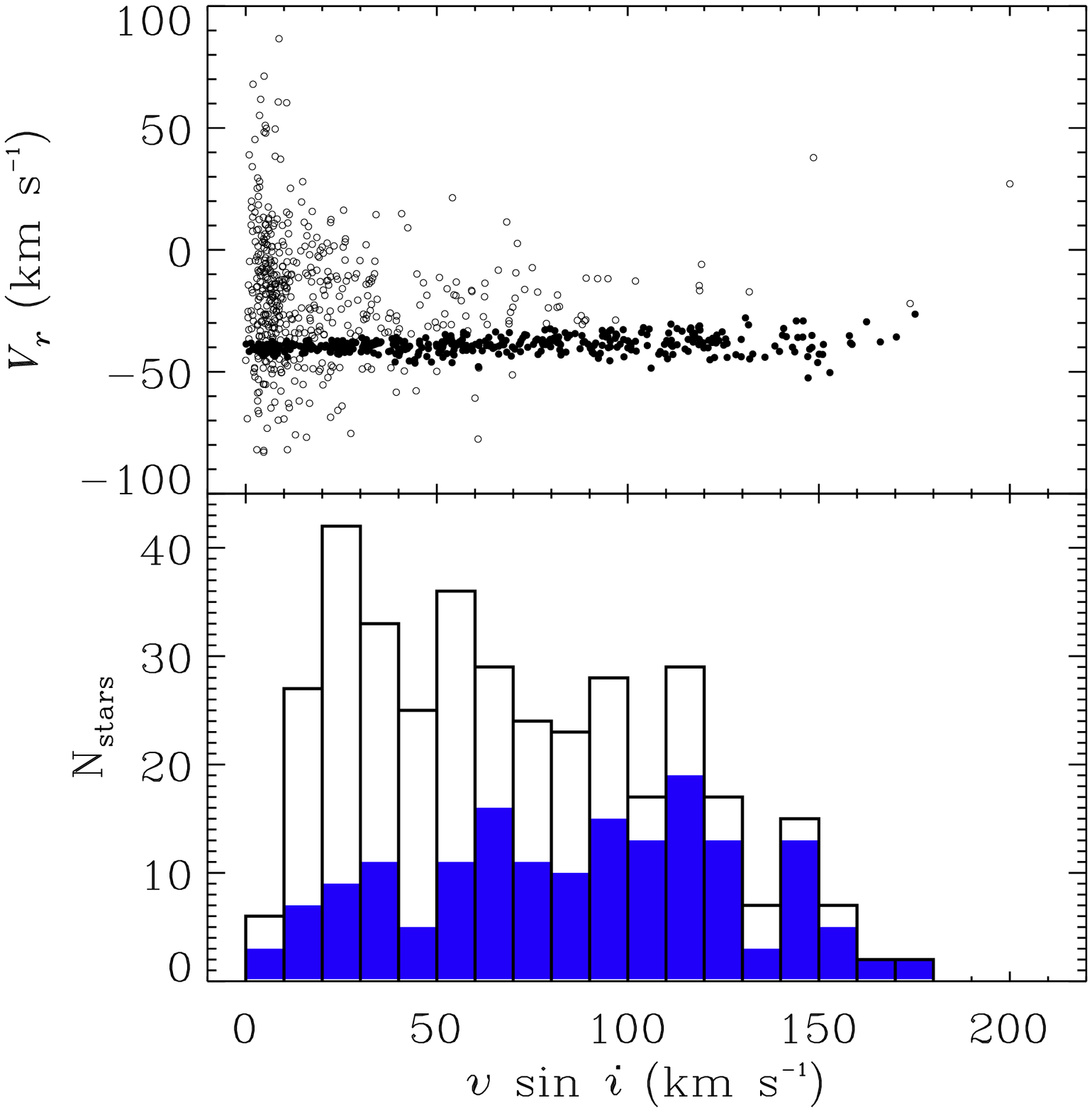}
\caption{Upper panel: radial velocities as a function of $v \sin i$.
The large dots indicate our selection of probable kinematic cluster
members of Tr~20. The widening range of cluster $V_{\rm r}$ is due
to the growing uncertainties at higher $v \sin i$.
Lower panel: a histogram of the main-sequence probable
cluster members.  A sample used to examine the effects of
stellar rotation (see Section 4) is marked by shaded area.
\label{fig1}}
\end{figure}

\clearpage

\begin{figure}
\epsscale{.95}
\plotone{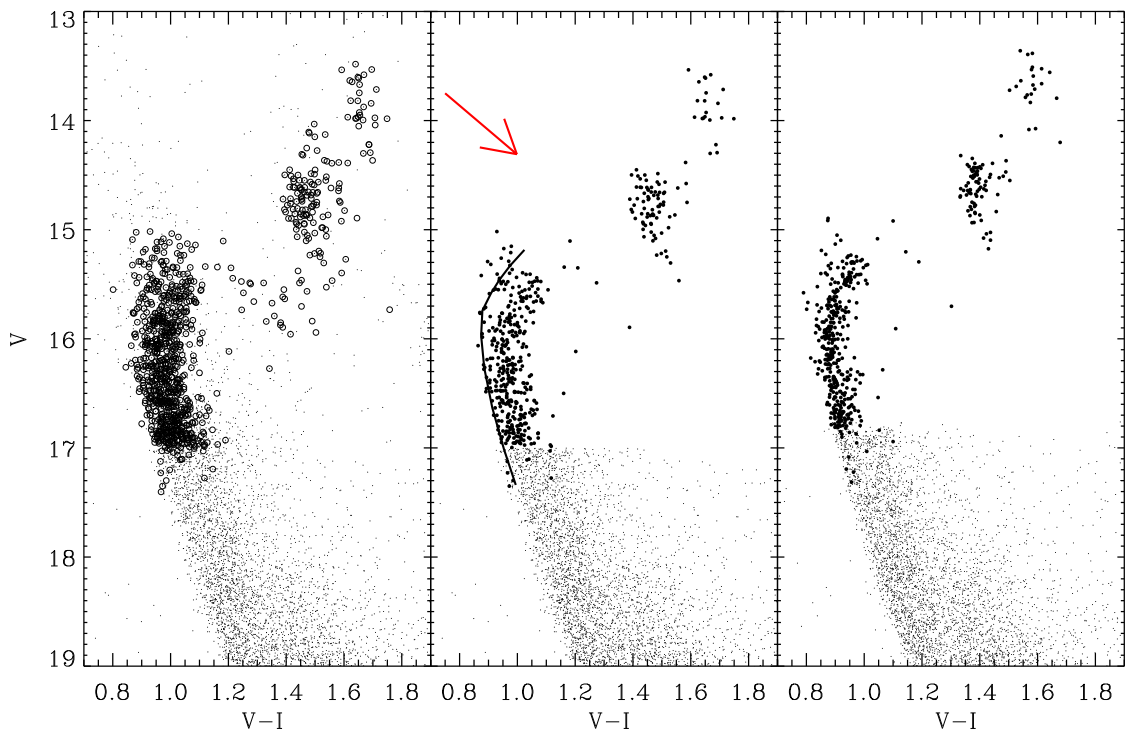}
\caption{Color--magnitude diagram of Tr~20. Left panel: the small dots
show all stars within $r < 8\arcmin$ from the cluster center. The
circles indicate our selection of targets for VLT/FLAMES spectroscopy.
Middle panel: the larger dots are $V_{\rm r}$-selected probable cluster
members. The arrow shows the direction and a half of the adopted amount of
reddening, $E(V-I)=0.50$. The location of the upper main sequence
at this lowest reddening is indicated by the curve. The presumable field stars
are deleted at $V<17$. Right panel: all stars shown in the middle
panel but corrected for differential reddening. 
\label{fig2}}
\end{figure}

\clearpage

\begin{figure}
\epsscale{.55}
\plotone{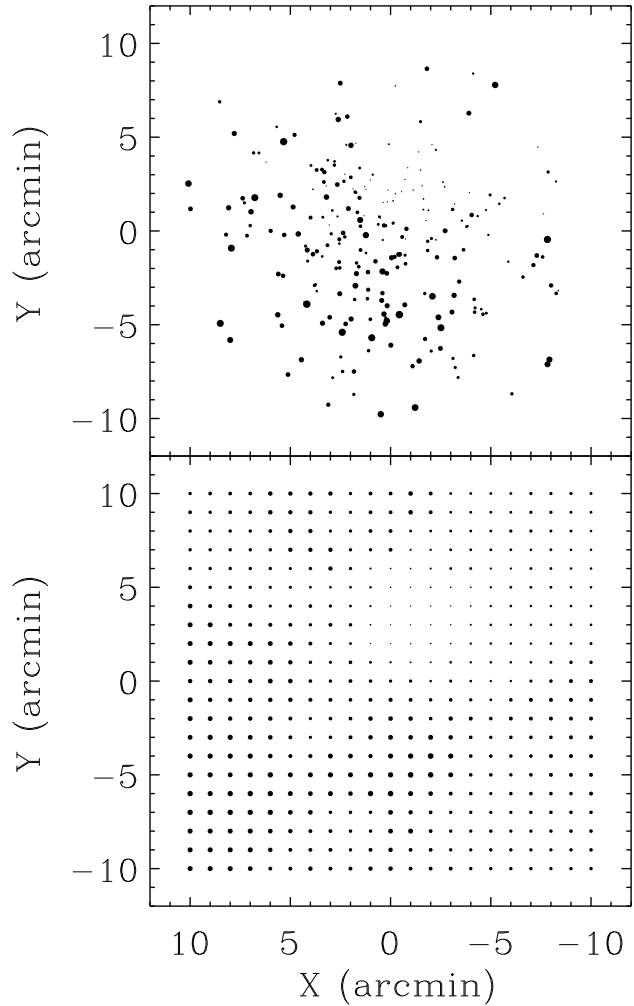}
\caption{Spatial distribution of differential reddening (DR) in Tr~20.
Upper panel: estimated DR using the main-sequence stars with $v \sin i < 95$
km~s$^{-1}$. The size of filled circles is proportional to the amount
of DR, ranging from 0.0 to $\sim$0.13~mag in $E(V-I)$.
A window of lower reddening is clearly visible at $X\sim-1\arcmin$,
$Y\sim+3\arcmin$. Lower panel: a grid of smoothed DR. In the corners of
this grid the estimates of DR are extrapolated and uncertain.
\label{fig3}}
\end{figure}

\clearpage

\begin{figure}
\epsscale{.95}
\plotone{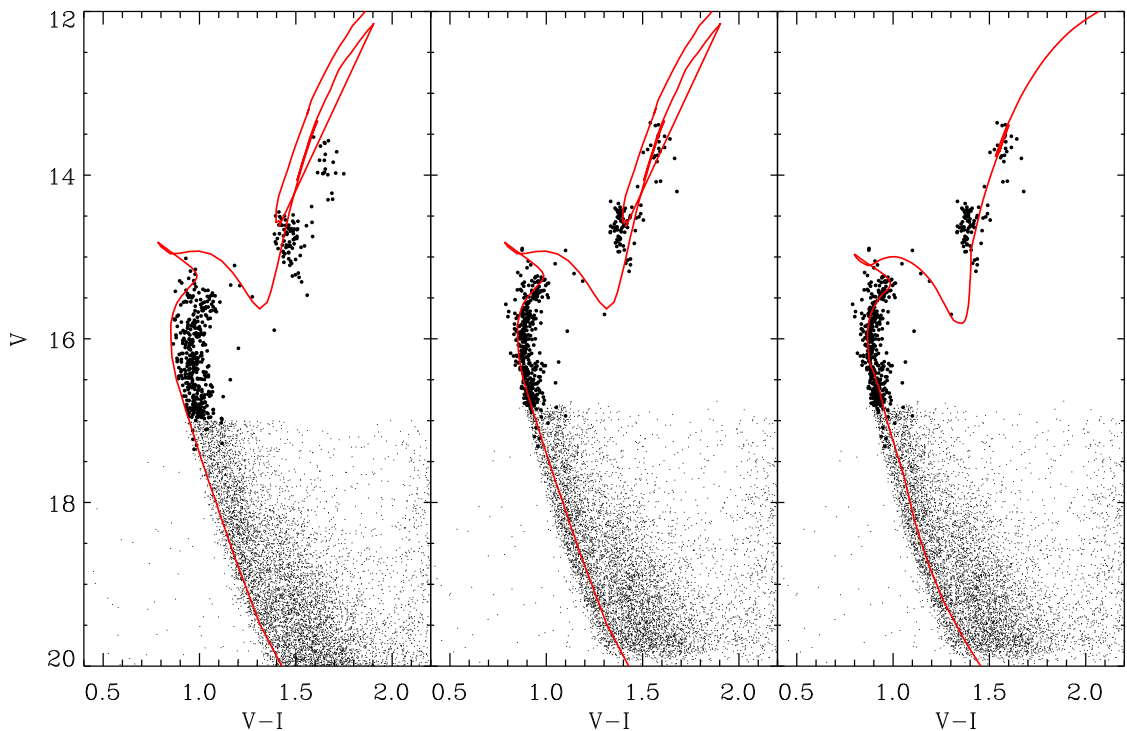}
\caption{Color--magnitude diagram of Tr~20 and a fit with theoretical
isochrones. Left panel: same as Figure~2 (middle panel). Provisional
fit with a Padova 1.3~Gyr isochrone targets the cluster's ``blue''
main sequence. A poor fit to the features of
RGB is evident. Middle panel: same as Figure~2 (right panel) and the
overplotted Padova 1.3~Gyr isochrone.
The tightness of a fit has improved substantially. There is 
evidence that the morphology of the red clump is complex.
Right panel: a fit with a Dartmouth 1.5~Gyr isochrone.
The reasoning of an older age is given in the text. In all three
cases the fit parameters (see Section 3.2) are kept unchanged.
\label{fig4}}
\end{figure}

\clearpage

\begin{figure}
\epsscale{.75}
\plotone{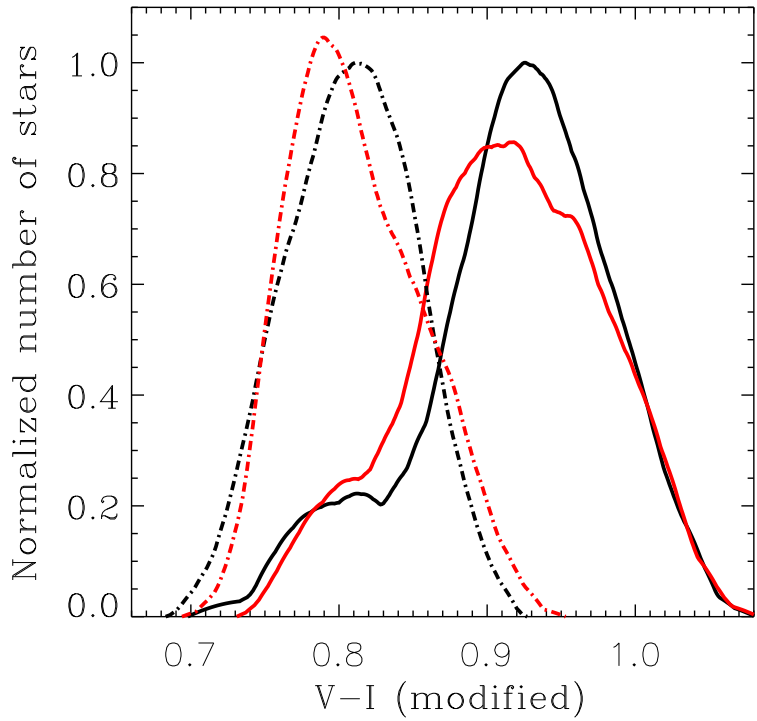}
\caption{Color distribution of the upper main-sequence stars.
The sample of 168 stars with $V<16.3$
is divided at $v \sin i = 90.5$ km~s$^{-1}$ in two equal halves of slow
(dark curve) and fast (red curve) rotators. The dotted curves
represent the normalized and smoothed histograms after correcting the
abscissa for differential reddening, while continuous curves are
constructed with uncorrected data. In both cases, the fast rotators
are slightly blueshifted. 
\label{fig5}}
\end{figure}

\end{document}